%\nopagenumbers
\magnification 1260 \baselineskip=14.8pt \voffset=-0.2truecm
\hoffset=-1.1truecm \vsize=24.8truecm \hsize=17.8truecm
\font\gf=cmb10 scaled 1300  \null \vskip
1.0truecm \centerline{\gf Binding energies of hydrogen-like
impurities in a} \vskip 0.1truecm\centerline{\gf semiconductor in
intense terahertz laser fields} \vskip 0.8truecm \centerline{J.L.
Nie$^{1,2)}$, W. Xu$^{1),*}$, and L.B. Lin$^{2)}$} \vskip
0.5truecm \centerline{$^{1)}$ Department of Theoretical Physics}
\vskip 0.1truecm \centerline{Research School of Physical Sciences
and Engineering} \vskip 0.1truecm \centerline{Australian National
University, Canberra ACT 0200, Australia} \vskip 0.1truecm
\centerline{$^{2)}$ Department of Physics, Sichuan University}
\vskip 0.1truecm \centerline{Chengdu - 610064, The People's
Republic of China}


\vskip1.0truecm We present a detailed theoretical study of the
influence of linearly polarised intense terahertz (THz) laser
radiation on energy states of hydrogen-like impurities in
semiconductors. The dependence of the binding energy for ground
($1s$) and first excited ($2s$) states, $E_{1s}$ and $E_{2s}$, on
intensity and frequency of the THz radiation has been examined for
a GaAs-based system. It is found that $E_{1s}$, $E_{2s}$ and
$E_{2s}-E_{1s}$ decrease with increasing radiation intensity or
with decreasing radiation frequency, which implies that an intense
THz field can enhance ionisation of dopants in semiconductors. Our
analytical and numerical results show that one of the most
important results obtained by A.L.A. Fonseca {\it et al} [phys.
stat. sol. (b) {\bf 186}, K57 (1994)] is incorrect.

\vfill\eject \leftline{\bf 1. Introduction} \vskip 0.3truecm The
ionisation of dopants (i.e., donor and acceptor impurities)
provides the major source of carriers (i.e., electrons and holes)
in semiconductors. The ionised dopants are also the major sources
of electron-impurity scattering in semiconductor devices, which
determine the transport and optical properties of the device
systems at low-temperatures. Hence, the investigation of
ionisation of dopants (such as binding energies of donors and
acceptors, transition energies and probabilities among different
impurity states, etc.) is fundamental in understanding almost all
physically measurable properties in semiconductors. In the absence
of an intense electro-magnetic (EM) radiation field, the binding
energies of dopants and the transition energies among different
impurity states in popularly used semiconductor materials are
known [1] and the theoretical approaches to calculate these
impurity states are well-documented [2].

It should be noted that in semiconductor materials such as GaAs,
Ge and Si, the binding energies of donor and acceptor impurities
are of the order of terahertz ($10^{12}$ Hz or THz) photon
energies [3] so that an intense THz radiation can affect strongly
the impurity states. With development and application of coherent,
high-power, long-wavelength, frequency-tunable and linearly
polarised radiation sources such as THz or far-infrared (FIR)
free-electron lasers (FELs) [4], it has now become possible to
measure the effect of an intense laser radiation on ionisation and
perturbation of dopants (especially shallow impurities) in
different semiconductor systems. In recent years, using THz FELs
[such as Free Electron Laser for Infrared eXperiments (FELIX) in
The Netherlands and CW FELs at UCSB] as intense radiation sources,
THz-photon-induced impact ionisation in InAs heterostructures [5],
time-resolved shallow donor spectrum in Si-doped GaAs [6] and
Lyman transitions in Be-doped GaAs [7] have been investigated
through, e.g., transport and/or photoconduction measurements. The
results obtained experimentally indicate that in the presence of
intense laser radiation such as FEL fields, i) impurity states in
different semiconductor systems are perturbated by the intensity
and frequency of the THz laser fields [5-7]; ii) photoconduction
experiments are more sensitive than optical measurements for
detecting transition energies of impurity states in semiconductors
[6,7]; and iii) some interesting intense radiation phenomena, such
as impact ionisation of dopants [5] and splitting and broadening
of the impurity spectrum [7], can be observed. In order to
understand these fundamentally new experimental findings and to
predicate new radiation phenomena, it is essential to know
theoretically how an intense laser field affects the binding
energies of impurities in semiconductors and is the prime
motivation of the present study.

In this paper, we study binding energy of hydrogen-like impurities
in bulk semiconductors in the presence of intense THz laser
fields. For semiconductor materials such as GaAs, there is an
absolute conduction band minimum at ${\bf K}=0$ with $\Gamma_6$
symmetry and the electron-effective-mass and dielectric constant
can be considered to be isotropic. Therefore, the impurity states
in materials such as GaAs are strictly scaled-down versions of
those in the hydrogen atom. The small electron-effective-mass
coupled with the large static dielectric constant result in small
binding energy for donors in these materials [3]. It has been
noticed by atomic physicists in their early work that intense UV
radiation can affect strongly the structure of atomic hydrogen
[8]. Using theoretical approaches proposed and developed in Ref.
[8], Nunes and co-workers have investigated the influence of
intense laser radiation on hydrogen-like impurity states in bulk
semiconductors [9] and in semiconductor quantum well structures
[10]. However, in Ref. [9], only the ground-state energy at
high-intensity radiation was presented analytically. We find that
it is hard to understand the result given by Eq. (13) in Ref. [9],
because it shows that at high-intensity radiation limit (ie,
$\alpha_0\to\infty$) the ground-state binding energy $E_0\to
-R_y^*$ (binding energy in the absence of the radiation) with
$R_y^*$ being the effective Rydberg constant. In order to see how
impurity states in a semiconductor are perturbated within the
frequency and intensity range of the current generation of the THz
FELs, we think it is necessary to re-examine the dependence of the
impurity binding energy in a semiconductor on frequency and
intensity of the THz fields. Moreover, since most of the
experimental work in this area has been carried out using
photoconduction experiments [6,7] which measure the transition
energies among different impurity states, it is also necessary to
know how an intense THz radiation affects the excited impurity
states in a semiconductor. In Section 2, we will first briefly
introduce the theoretical approaches used for the calculations
then present analytical results of the binding energies for ground
and first excited impurity states. The numerical results will be
presented and discussed in Section 3 and the conclusions drew from
this study will be summarised in Section 4.

\vskip 0.5truecm \leftline{\bf 2. Analytical results} \vskip
0.3truecm

In this work, we consider the situation where a laser field with a
vector potential $A(t)$ is applied along the x-direction of a
semiconductor and the radiation field is polarised linearly along
the z-direction. Under the effective mass approximation, the
time-dependent Hamiltonian to describe the electron-impurity
system is given as $$H(t)={1\over
2m^*}[p_x^2+p_y^2+(p_z-eA(t))^2]-{e^2\over\kappa R}.\eqno(1)$$
Here, $m^*$ is the electron-effective-mass, $p_x=-i\hbar\partial \
/\partial x$ is the momentum-operator along the x-direction, under
the dipole approximation $A(t)=(F_0/\omega){\rm sin}(\omega t)$ is
the vector potential induced by the radiation field with $\omega$
being the radiation frequency and $F_0$ the electric field
strength of the radiation field, ${\bf R}=({\bf r},z)=(x,y,z)$,
and $e^2/\kappa R$ is the Couloumb potential induced by
electron-impurity interaction with $\kappa$ being the dielectric
constant. It can be seen that due to time-dependent nature of the
radiation field, in principle, we have to solve time-dependent
Schr\"odinger equation: $[i\hbar\partial / \partial t - H(t)]\Psi
({\bf R},t)=0 $ to obtain the electronically bounded impurity
states. However, at present, there is no simple analytical
solution to this problem [11]. Here we employ a tractable
theoretical approach proposed and developed by atomic physicists
[8] to study the binding energy of the impurity states. The key
issue of this approach is to find laser-dressed potential energy.
It is found that if the coordinate along the z-axis is shifted by
$z'=z+z_0(t)$ with $z_0(t)=-(eF_0/m^*\omega^2){\rm cos}(\omega
t)$, the Schr\"odinger equation becomes $$i\hbar{\partial
\Psi({\bf R'},t)\over\partial t}=\Bigl[{p_x^2 +p_y^2+p_{z'}^2
\over 2m^*}+V({\bf R'},t) \Bigr]\Psi({\bf R'},t), \eqno(2a)$$
where ${\bf R'}=({\bf r},z')$ and $$V({\bf R'},t)=-{e^2\over
\kappa\sqrt{r^2+[z'-z_0 (t)]^2}} +{[eA(t)]^2\over
2m^*}.\eqno(2b)$$ After averaging $V({\bf R'},t)$ over a period of
the radiation field, we obtain laser-dressed potential as $$V({\bf
R'})={\omega \over 2\pi} \int_0^{2\pi/\omega} dt\ V({\bf
R'},t)=E_{em}-{e^2\over 2\kappa}[|{\bf R'}+{\bf R}_0|^{-1} +|{\bf
R'}-{\bf R}_0|^{-1}].\eqno(3)$$ Here
$E_{em}=(eF_0)^2/4m^*\omega^2$ is the energy shift induced by the
radiation field due to dynamical Franz-Keldysh effect [12] and
${\bf R}_0=(0,0,eF_0/m^*\omega^2)$. It should be noted that in
atomic physics, $E_{em}$ is also called ponderomotive energy [13].
If we replace $V({\bf R}',t)$ in Eq. (2a) by $V({\bf R'})$ given
by Eq. (3), the time-dependent problem becomes approximately a
time-independent one and the solution is $\Psi({\bf
R}',t)=e^{-i(E_N+E_{em})t/\hbar}\Psi_N ({\bf R}')$, where the
wavefunction $\Psi_N ({\bf R}')$ and the energy spectrum $E_N$ are
determined by $(H_0-E_N)\Psi_N({\bf R}')=0$ with
$$H_0={p_x^2+p_y^2+p_{z'}^2\over 2m^*}-{e^2\over 2\kappa} (|{\bf
R}'+{\bf R}_0|^{-1}+|{\bf R}'-{\bf R}_0|^{-1}).\eqno(4)$$

In contrast to Refs. [8-10], in the derivations shown above we
have included the terms associated with $F_0^2$ which, as can be
seen, only contributes to the energy shift induced by the
radiation field alone. The main merit of this approach is that now
one can convert a time-dependent problem into a time-independent
one and, thus, one can use, e.g., variational method to calculate
the energy states and the corresponding wavefunctions. It should
be noted that the present theoretical approach is a generalization
of those proposed by the atomic physicists. As has been pointed
out by Ref. [13], this approach is the lowest order approximation
to the atomic problem. Furthermore, the results given by Eqs. (3)
and (4) have been documented by Refs. [9,10] for the case of the
hydrogen-like impurities in a semiconductor.

In the present work, we employ the wavefunctions for a hydrogen
atom as the trial functions for ground state (ie, $1s$-state) and
first excited state (ie, $2s$-state) of the donor impurities,
which reads
$$\Psi_{1s}({\bf R}')=a e^{-\beta R'/a_0^*} \eqno(5a) $$ and
$$ \Psi_{2s}({\bf R}')=c(1+b R'/a_0^*)e^{-\gamma
R'/2a_0^*},\eqno(5b)$$ where $\beta$ and $\gamma$ are variational
parameters, $a_0^*=\hbar^2\kappa/m^*e^2$ is the effective Bohr
radius, $a=(\beta^3/\pi {a_0^*}^3)^{1/2}$ is determined by
$<1s|1s>=1$, and $b=-(\beta+\gamma/2)/3$ and $c=\gamma^{5/2}/[8\pi
{a_0^*}^3 (12b^2+6b\gamma+\gamma^2)]^{1/2}$ are determined
respectively by $<1s|2s>=0$ and $<2s|2s>=1$. After introducing
these trial functions to the variational method, i.e.,
$E_N=<N|H_0|N>$, the binding energies of the $1s$-state and
$2s$-state are obtained as $$E_{1s}=-R_y^*
\Bigl[{2\over\alpha_0^*}-\beta^2- {2\over \alpha_0^*}(1+\alpha_0^*
\beta) e^{-2\alpha_0^* \beta} \Bigr],\eqno(6a)$$ and
$$E_{2s}=-{R_y^*\over 4} {(4/\alpha_0^*)(24b^2+12b\gamma
+2\gamma^2-{\cal A} e^{-\alpha_0^*\gamma})-
\gamma^2(4b^2+2b\gamma+\gamma^2)\over 12b^2+6b\gamma+
\gamma^2},\eqno(6b)$$ where $\alpha_0^*=eF_0/m^*\omega^2 a_0^*$,
$R_y^*=m^*e^4/2\hbar^2 \kappa^2$ is the effective Rydberg
constant, and $${\cal
A}=24b^2+12b\gamma+2\gamma^2+\alpha_0^*\gamma
(18b^2+8b\gamma+\gamma^2)+2{\alpha_0^*}^2 b\gamma^2 (3b+
\gamma)+{\alpha_0^*}^3 b^2\gamma^3.$$ Furthermore, the variational
parameters $\beta$ and $\gamma$ are determined respectively by
$$\beta-(1+2\alpha_0^* \beta) e^{-2\alpha_0^* \beta}=0 \ \ \ \ \
{\rm and} \ \ \ \ \ \partial E_{2s}/ \partial \gamma=0.$$

The results shown above indicate that the effect of the radiation
field on binding energy of the impurities is mainly achieved
through a dimensionless factor
$\alpha_0^*=e^3F_0/\hbar^2\kappa\omega^2\sim F_0/\omega^2$. For
high-frequency and/or low-intensity radiation so that
$\alpha_0^*\ll 1$, we have $$E_{1s}\simeq -R_y^* \Bigl(1-{4\over
3}{\alpha_0^*}^2\Bigr) \ \ \ \ {\rm and} \ \ \ \ E_{2s}\simeq
-{R_y^*\over 4}\Bigl(1-{2\over 3}{\alpha_0^*}^2 \Bigr), \eqno(7)$$
which suggests that the binding energy decreases quickly with
increasing radiation intensity or with decreasing radiation
frequency when $\alpha_0^*\sim 1$. When $\alpha_0^*=0$,
$E_{1s}=-R_y^*$ and $E_{2s}=-R_y^*/4$ are well-known results
obtained in the absence of the radiation field. For low-frequency
and/or high-intensity radiation, entailing $\alpha_0^*\gg 1$, we
find that there is no simple analytical expression for both
$E_{1s}$ and $E_{2s}$, in contrast to Eq. (13) obtained by Ref.
[9]. When $\alpha_0^*\to \infty$, we find $\beta\to 0$ and
$\gamma\to 0$ so that $E_{1s}\to 0$ and $E_{2s}\to 0$ (note that a
zero binding energy implies that all impurities at this state are
ionised). These theoretical results indicate that $E_{1s}$ and
$E_{2s}$ are altered from respectively $-R_y^*$ and $-R_y^*/4$ at
zero field to zero at high-field limit. One of the most important
results obtained by Ref. [9] is that at high-intensity radiation
fields, the ground-state binding energy is given by Eq. (13) which
shows $E_{1s}\to -R_y^*$ when $\alpha_0\to \infty$ and $E_{1s}
<-R_y^*$. Our results here suggest that Eq. (13) obtained by Ref.
[9] is incorrect.

\vskip 0.5truecm \leftline{\bf 3. Numerical results} \vskip
0.3truecm Below we present numerical results for semiconductor
materials such as GaAs. The material parameters for GaAs taken
within the calculations are the effective-electron-mass ratio
$m^*/m_e=0.0665$ with $m_e$ being the electron rest mass and the
static dielectric constant $\kappa=12.9$. The dependence of
binding energies $E_{1s}$ and $E_{2s}$ as well as transition
energy $E_{2s}-E_{1s}$ on THz laser radiation fields is shown in
Figures 1 - 4. From these results, we see that:

\item{a)} in the presence of the radiation fields, $E_{1s}$ and
$E_{2s}$ are altered from respectively $-R_y^*$ and $-R_y^*/4$ at
low-field limit to zero at high-field limit and $E_{2s}-E_{1s}$ is
altered from $3R_y^*/4$ at low-field limit to zero at high-field
limit. This confirms that Eq. (13) obtained by Ref. [9] is
incorrect;

\item{b)} with increasing radiation intensity and/or decreasing
radiation frequency, $E_{1s}$, $E_{2s}$ and $E_{2s}-E_{1s}$
decrease;

\item{c)} $E_{2s}$ depends a bit weakly on the radiation field
than $E_{1s}$ does, which can be understood by Eq. (7);

\item{d)} the strong effect of the radiation field on binding
energy and transition energy can be observed at $\alpha_0^*\sim
1$; and

\item{e)} $E_{1s}$, $E_{2s}$ and $E_{2s}-E_{1s}$ depend more
strongly on radiation frequency than on radiation intensity
because $\alpha_0^*\sim F_0/\omega^2$.

The results discussed above indicate that in the presence of the
intense THz laser fields, the binding energies of the impurity
states can be reduced and the impurity spectrum in semiconductors
can be shifted significantly by the radiation.

It should be noted that the current generation of the FELs can
provide intense THz radiation sources in the frequency and
intensity range $f\sim 0.1\ -\ 10$ THz and $F_0\sim 0.1 \ -\ 100$
kV/cm [14] so that the condition $\alpha_0^*=e^3F_0/\hbar^2
\kappa\omega^2\sim 1$ can be satisfied by most of the popularly
used semiconductor materials. We therefore believe that it has now
become possible to investigate the effects of the intense laser
radiation on ionisation and perturbation of dopants in
semiconductor systems by using current generation of THz or FIR
FELs.

\vskip 0.5truecm \leftline{\bf 4. Further remarks}\vskip 0.3truecm

In the present study, our calculations have been performed for
shallow-impurities in semiconductor materials such as GaAs,
because these impurities are hydrogen-like. It is well known that
in semiconductors, deep-impurities may not be hydrogen-like
because their states depend strongly on lattice structure of the
material systems, such as the symmetry of the crystal potential,
electronic band structure, etc. To examine the effect of intense
laser radiation on deep-impurity states in a semiconductor, the
above mentioned theoretical approach has to be modified.

To our knowledge, except those theoretical results reported by
Refs. [9] and [10] for semiconductor systems, so far very few
theoretical and experimental results have been reported regarding
the {\it direct} effects of the intense laser radiation on binding
energies of electrons in atomic systems (two recent reviews are
given in Refs. [13] and [15]) or of hydrogen-like impurities in
semiconductor systems. The main reason behind this is that when an
electronic system (atoms or semiconductors) is subjected to
intense laser fields, the investigation of the electronic states
(or electronically bounded impurity states) is essentially a
time-dependent problem. As a consequence, these states are no
longer the eigenstates and transitions among these time-dependent
states can occur. Therefore, it is very hard to measure
experimentally the binding energy for a certain atomic or impurity
state when an intense laser radiation is present. Normally, the
results obtained from atomic physics experiments [13,15] are
contributions from all possible states and transitions among these
states.

The analytical and numerical results shown in this paper are what
one can get from the simplest theoretical model. As has been
pointed out by Ref. [13], the approach used in the present study
is the lowest order approximation to deal with the problem. Under
such an approximation, the systems can be described in terms of a
set of quasi-eigenstates of atom (or impurity) and laser field and
there is no transition among these quasi-eigenstates. Namely,
under this approximation, the system is ``stable to ionisation''.
Thus, the results obtained from this model can be useful in
understanding atomic physics phenomena such as free-free
transitions and dichotomy of the atomic states [8]. Furthermore,
it can be shown [13] that the conditions under which the approach
can be applied are $\hbar\omega \gg E_{em}$ and $R_0^2\omega \gg
1$. Therefore, the very low-frequency and high-intensity results
shown in Figs. 1-4 in this paper may not be the case. To examine
the effect of these extreme radiation conditions on states of
hydrogen-like impurities in a semiconductor, the contributions
from higher order terms (see Eq. (28) in Ref. [13]) have to be
included. The inclusion of these higher order contributions
requires considerably further analytical and numerical
calculations and, therefore, we do not attempt it in the present
study.

\vskip 0.5truecm \leftline{\bf 5. Conclusions}\vskip 0.3truecm In
this paper, we have re-examined how a linearly polarised intense
laser field affects the binding and transition energies of
shallow-donor impurities in a bulk semiconductor. The present
study has been conducted on the basis of a theoretical approach
proposed by Ref. [8] and documented by Refs. [9,10]. We have
obtained results of binding energies for both ground and first
excited impurity states and thus we can look into the effects of
the radiation field on transition energy between $1s$ and $2s$
states, in conjunction with experimental work carried out by
photoconductivity measurements [6,7]. The numerical and analytical
results obtained from this work show that Eq. (13) in Ref. [9] is
incorrect.

We have found that the strong influence of the radiation field
on binding and transition energies of the shallow impurity
states in bulk semiconductors can be observed when the
condition $\alpha_0^*=e^3F_0/\hbar^2\kappa\omega^2\sim 1$ is
satisfied. For most popularly used semiconductor materials
such as GaAs, Si and Ge,
these radiation conditions have been realised by the current
generation of the THz or FIR FELs. We therefore hope that the
phenomena studied and predicated in this paper can be verified
experimentally.

\vskip 0.5truecm \leftline{\bf Acknowledgment}\vskip 0.5truecm
One of us (W.X.) is a Research Fellow of the Australian Research
Council (ARC). This work was also supported by the ARC IREX Grant,
Visiting Scholar Foundation of Key Laboratory in University of
the Ministry of Education, China and National Natural Science
Foundation of China. Discussions with P. Fisher and R.A. Lewis
(UoW, Australia) are gratefully acknowledged.

\vfill\eject \leftline{\bf References} \vskip 0.5truecm
\item{$^*$} Corresponding author and electronic mail:
wen105@rsphysse.anu.edu.au

\item{[1]} For a review, see, e.g., A.K. Ramdas and S. Rodriguez,
Rep. Prog. Phys. {\bf 44}, 1297 (1981).

\item{[2]} For a early work, see, e.g., J.M. Luttinger and W.
Kohn, Phys. Rev. {\bf 97}, 869 (1955).

\item{[3]} For example, in the absence of a radiation field, the
donor binding energy in n-GaAs is 5.4 meV (or 1.3 THz) and for P
in Si is 45.6 meV (or 11 THz).

\item{[4]} For recent development of the FELs, see, eg, {\it Free
Electron Lasers 1999}, edited by J. Feldhaus and H. Weise
(Elsevier Science, North Holland, 2000).

\item{[5]} A.G. Markelz, N.G. Asmar, B. Brar and E.G. Gwinn, Appl.
Phys. Lett. {\bf 69}, 3975 (1996).

\item{[6]} P.C.M. Planken, P.C. van Son, J.N. Hovenier, T.O.
Klaassen, W. Th. Wenckebach, B.N. Murdin and G.M.H. Knippels,
Phys. Rev. {\bf B 51}, 9643 (1995); Infrared Phys. Technol. {\bf
36}, 333 (1995).

\item{[7]} R.A. Lewis, L.V. Bradley, and M. Henini, Solid State
Commun. {\bf 122}, 223 (2002).

\item{[8]} M. Gavrila and J.Z. Kaminski, Phys. Rev. Lett. {\bf
52}, 613 (1984); M. Point, N.R. Walet, M. Gavrila, and J. McCurdy,
Phys. Rev. Lett. {\bf 61}, 939 (1988).

\item{[9]} A.L.A. Fonseca, M.A. Amato, and O.A.C. Nunes, Phys.
Stat. Sol. (b) {\bf 186}, K57 (1994).

\item{[10]} Q. Fanyao, A.L.A. Fonseca, and O.A.C. Nunes, Phys.
Stat. Sol. (b) {\bf 197}, 349 (1996); Superlattices and
Microstructures {\bf 23}, 1005 (1998).

\item{[11]} A.D. Bandrauk and H.Z. Lu, Phys. Rev. {\bf A 62},
53406 (2002).

\item{[12]} W. Xu, Phys. Rev. {\bf B 57}, 15282 (1998).

\item{[13]} See, e.g., K. Burnett, V.C. Reed, and P.L. Knight, J.
Phys. B: At. Mol. Opt. Phys. {\bf 26}, 561 (1993).

\item{[14]} The connection between the electric field strength of
a laser field ($F_0$) and the laser output power ($I$) in vacuum
is: $I=0.5\sqrt{\epsilon/\mu} |F_0|^2\simeq 1.32 |F_0|^2$ kW
cm$^{-2}$ where $F_0$ is in units of kV/cm.

\item{[15]} For a recent review, see, e.g., M. Protopapas, C.H.
Keitel, and P.L. Knight, Rep. Prog. Phys. {\bf 60}, 389 (1997).

 \vfill\eject \leftline{\bf Figure captions} \vskip 0.5truecm
\noindent {\bf Fig. 1}: Binding energies, $E_{1s}$ and $E_{2s}$,
as a function of THz radiation frequency ($f=\omega/2\pi$) for
different radiation intensities. $R_y^*$ is the effective Rydberg
constant and for GaAs $R_y^*=5.44$ meV.

\vskip 0.3truecm \noindent {\bf Fig. 2}: Transition energy between
$2s$ and $1s$ impurity states, $E_{2s}-E_{1s}$, as a function of
radiation frequency for different radiation intensities.

\vskip 0.3truecm \noindent {\bf Fig. 3}: $E_{1s}$ and $E_{2s}$ as
a function of radiation intensity for different radiation
frequencies.

\vskip 0.3truecm \noindent {\bf Fig. 4}: $E_{2s}-E_{1s}$ as a
function of radiation intensity for different radiation
frequencies.

\vfill\eject \topinsert \input psfig.sty
\centerline{\psfig{figure=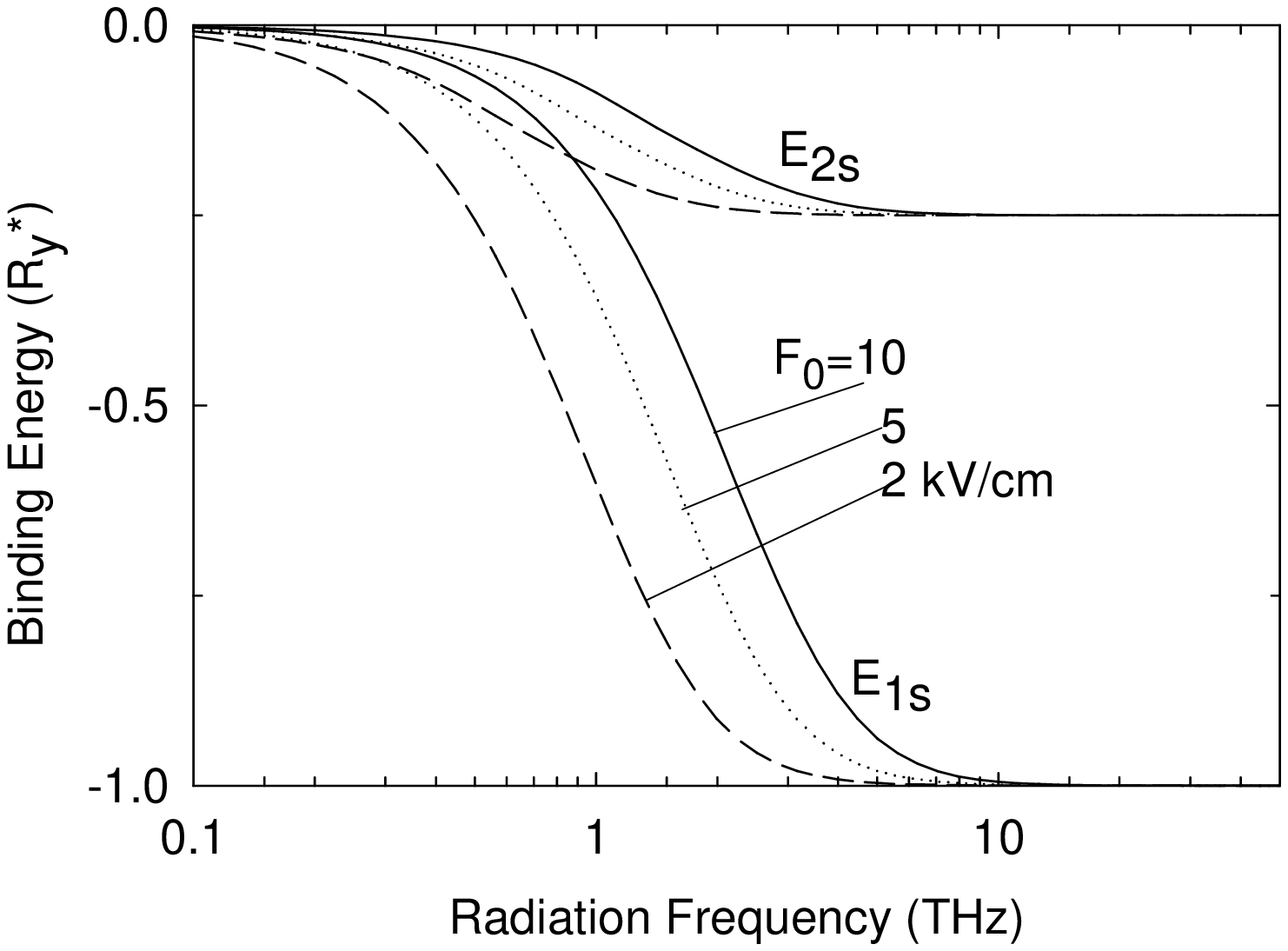,width=14truecm,angle=0}} \vskip
2.0truecm \noindent {\bf J.L. Nie et. al. Figure 1.} \endinsert

\vfill\eject \topinsert \input psfig.sty
\centerline{\psfig{figure=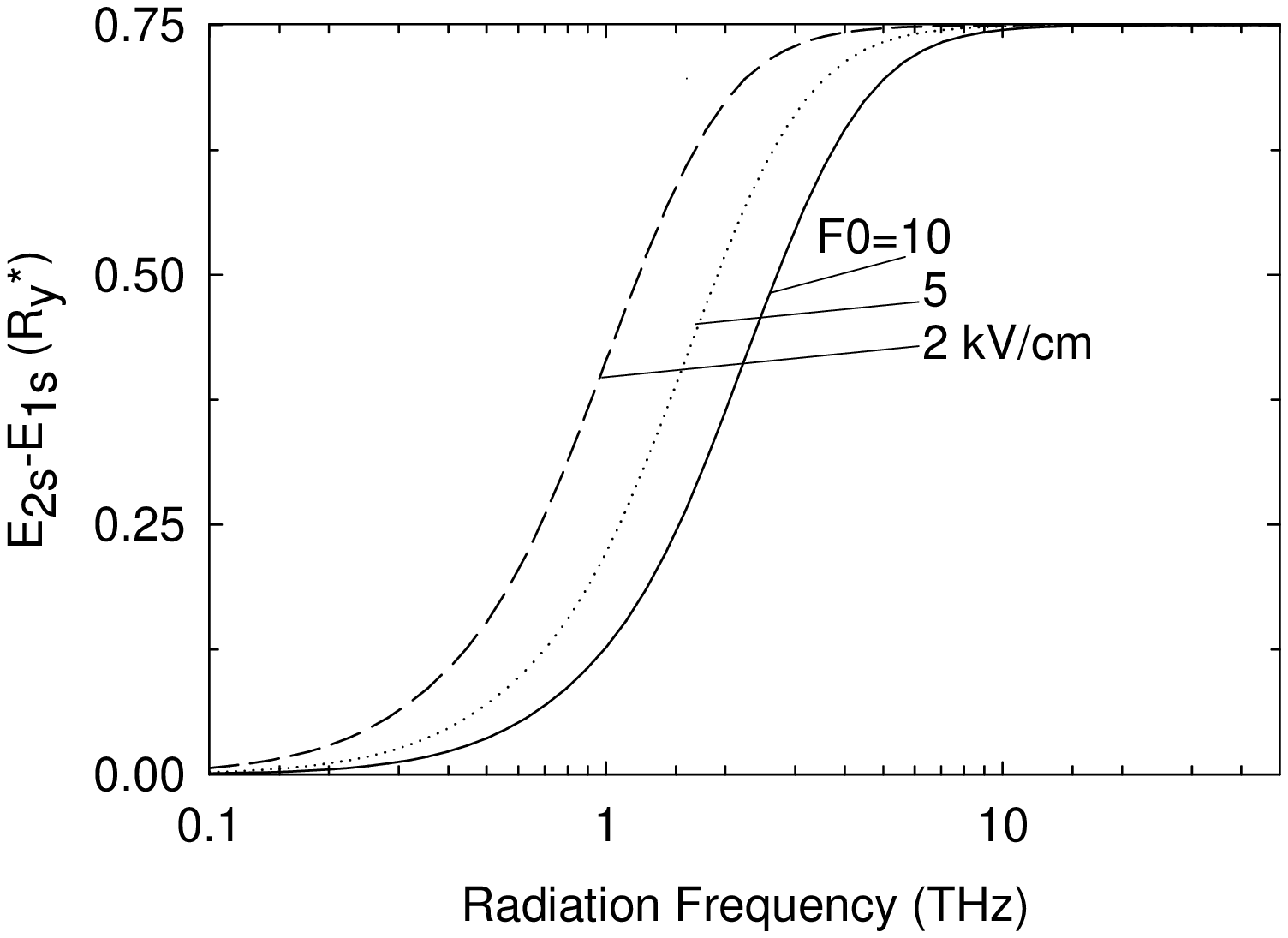,width=14truecm,angle=0}} \vskip
2.0truecm \noindent {\bf J.L. Nie et. al. Figure 2.} \endinsert

\vfill\eject \topinsert \input psfig.sty
\centerline{\psfig{figure=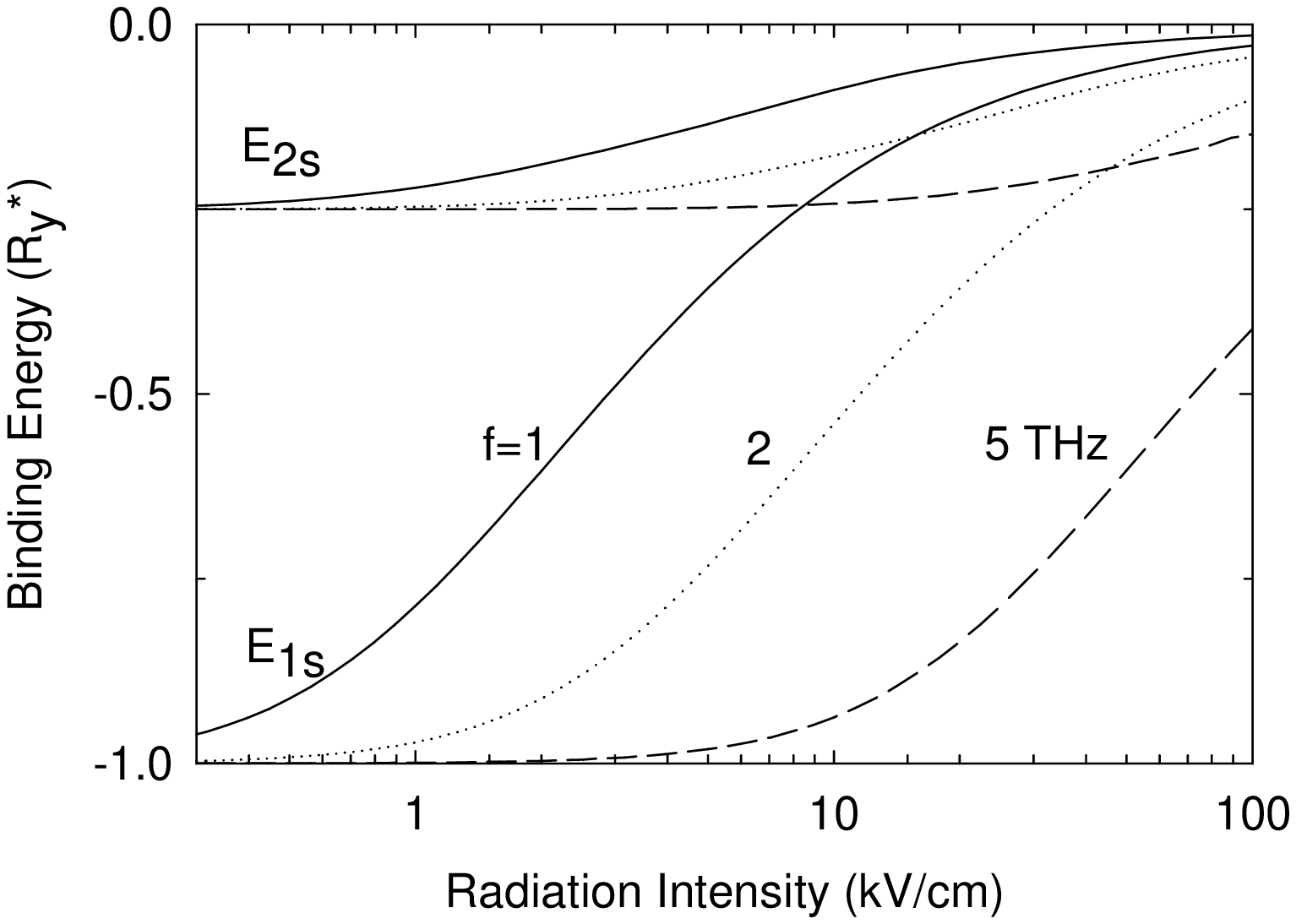,width=14truecm,angle=0}} \vskip
2.0truecm \noindent {\bf J.L. Nie et. al. Figure 3.} \endinsert

\vfill\eject \topinsert \input psfig.sty
\centerline{\psfig{figure=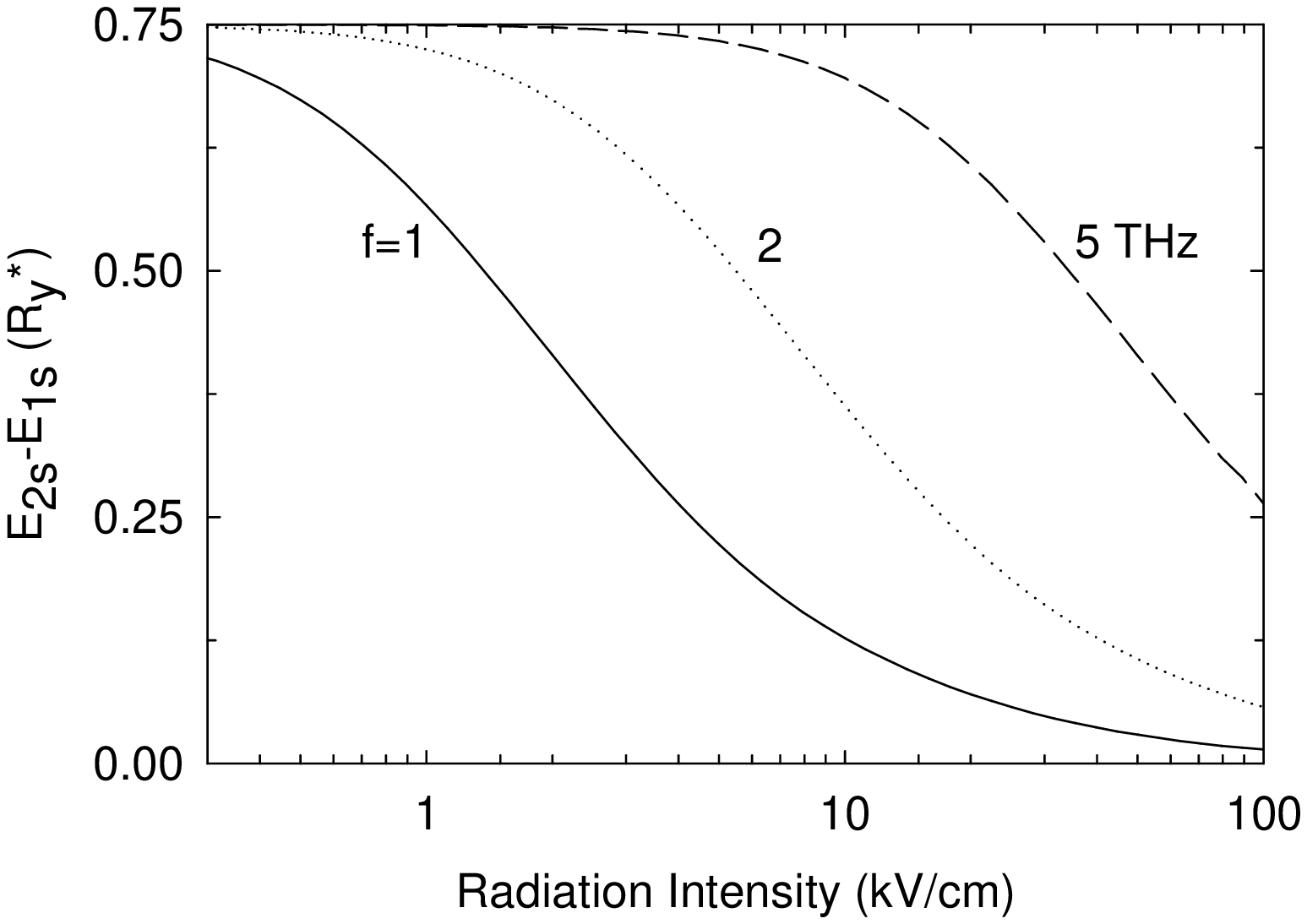,width=14truecm,angle=0}} \vskip
2.0truecm \noindent {\bf J.L. Nie et. al. Figure 4.} \endinsert
\end